\documentclass{JAC2000}
\usepackage{graphicx}

\begin{document}
\title{Energy stabilization of 1.5 GeV S-Band linac
%\thanks{Work supported by long suffering spouses and colleagues}
}

\author{M. Kuriki, H. Hayano, T. Naito, T. Okugi, KEK, Tsukuba, Japan\\
V. Vogel, BINP, Moscow, Russia}

\maketitle

\begin{abstract} 
KEK-ATF is studying low emittance, multi-bunch electron beam for the
future linear collider.  The energy instability of the 1.5 GeV linac
has been a problem making the beam injection to the damping ring
unstable.  Because the unstable beam generates also large amount of
the radiation, the beam current is limited by the KEK radiation safety
policy much lower than what we expect.  Stabilizing the S-band linac
is therefore important not only to improve the beam quality, but also
to clear the radiation safety limit to start the multi-bunch
operation.

We have made various modifications to solve the problem on the
electron gun, modulator, klystron etc. For the modulator, we have
developed a feed-forward controlled De-Q module. This module
compensates the voltage jitter by controlling the deQ timing with a
feed-forward circuit because the amount of the excessive charge up is
strongly correlated to the charge up slope that can be measured prior
to the deQ timing.  The energy stability was examined and was improved
by a factor of 3, from 0.6\% to 0.2\% of itself.  Modification for the
feed-forward circuit to get more stability was made. The test for the
new circuit is in progress.  For the long term instability, phase-lock
system for klystron RF is being installed.  In the test operation, it
showed a good performance and compensate the phase drift less than
$1^\circ$.

\end{abstract}

\section{Introduction}

KEK-ATF is a test facility to study the low emittance multi-bunch beam
and beam instrumentation technique for the future linear collider.
That consists from 1.5 GeV S-band linac, a beam transport line, a
damping ring, and a diagnostic extraction line.

In the linac, the electron beam is generated by a thermionic electron
gun. Typical intensity is $\rm 10^{10}$ electron/bunch. The bunch
length shrinks from 1 ns to 10 ps by passing a couple of sub-harmonic
bunchers, a TW buncher.  The electron beam is accelerated up to 1.3
GeV by 8 of the S-band regular accelerating section.  One section has
two accelerating structures driven by a klystron-modulator. Klystron
is Toshiba E3712 generating 80 MW with a pulse duration of $\rm 4.5
\mu s$ RF. A peak power of 400 MW with a pulse duration of $\rm 1.0\mu
s$ is obtained by SLED cavity and makes a high gradient accelerating
field, 30 MeV/m.

20 of bunches separated by 2.8 ns are accelerated by one RF pulse. This
multi-bunch method is one of the key technique in the linear collider,
but ATF is now operated in single-bunch mode because of the KEK
radiation safety policy.

In April 2000, we achieved horizontal emittance $\rm 1.3\times 10^{-9}
rad.m $, vertical emittance$1.7\times 10^{-11} rad.m $ (both for
$2.0\times 10^9 electron/bunch$, single bunch mode )\cite{emittance}
which are almost our target. 

So, our next task is to achieve such low emittance with the high
intensity, high reputation multi-bunch electron beam. To start the
multi-bunch operation, we have to clear the radiation safety limit anyhow.

One of the biggest source of the radiation is the beam transport
line. The large energy tail and the energy jitter causes the radiation
loss in the beam transport line.  To suppress the radiation ,  we have
to stabilize the beam energy and also improve the quality of the beam.

In addition, the energy jitter fluctuates the injection efficiency to
the damping ring, the beam intensity becomes therefore very unstable.
The energy stabilization of the linac is very important for the
systematic study for the beam instrumentation at the extraction line.

\section{Feed-forward deQ}
From a study of the energy instability\cite{hayano}, one of the
biggest source of the instability was klystron voltage jitter induced
by the fluctuated modulator output. We implement a new deQ method to
compensate it.

In our modulator, the pulse forming network, PFN is charged up
resonatorly up to 43 kV.  By discharging $\rm PFN$, a pulse
transformer gains this voltage up to 340 kV with pulse duration of 7.5
$\rm \mu s$ and is fed to the klystron.

\begin{figure}[htb]
\centering
\includegraphics*[width=50mm,height=45mm]{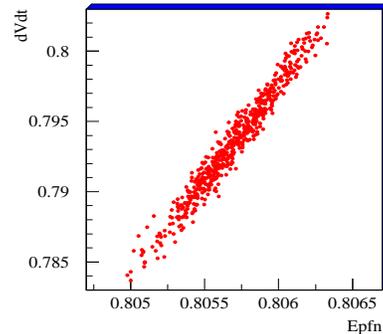}
%\epsfile{file=ffoff.eps,width=50mm,height=45mm}
\vspace{-5mm}
\caption{x and y axe show $\rm E_{PFN}$ and dV/dt respectively.  Even deQ works,
$\rm E_{PFN}$ correlates with dV/dt. }
\label{ffoff}
\end{figure}
deQ module controls the voltage of PFN, $\rm E_{PFN}$.  When $\rm
E_{PFN}$ reaches the reference voltage, deQ modules fires trigger to
stop the charge up. Fig.~\ref{ffoff} shows the correlation between
$\rm E_{PFN}$ and gradient of the charge up curve, $\rm dV/dt$.
Ideally $\rm E_{PFN}$ is independent from $\rm dV/dt$, but there is a
significant correlation.  This correlation is explained as follows;
there is some delay to stop charging up and the delay causes excessive
charge up which is expressed as $(dV/dt)\times \delta t$ where $\delta
t$ is the delay. This excess is therefore proportional to dV/dt and
induces the correlation.

If dV/dt was always same, it was not a problem.  In fact, the
amplitude of AC power is unstable and that makes fluctuated dV/dt.
Improving the quality of power source is straightforward, but it will
be rather expensive. Then we admit the fluctuation of dV/dt and
consider to compensate the $\rm E_{PFN}$ jitter.

\begin{figure}[htb]
\centering
\includegraphics*[width=80mm]{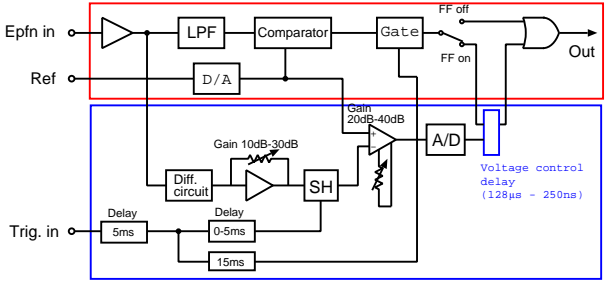}
%\epsfile{file=block.eps,width=80mm}
\vspace{-4mm}
\caption{Block diagram of the feed-forward controlled deQ.  Upper and lower parts are the conventional deQ and feed-forward circuit  respectively.}
\label{block}
\end{figure}

Because dV/dt is fluctuated and $\delta t$ could not be zero, then the
only way to compensate the fluctuation is to control $\delta t$ to
keep a constant excessive charge up. Fortunately, we can measure dV/dt
prior to fire deQ signal, then feed-forward control is possible.

\begin{table}[htbp]
\begin{center}
\caption{Stability of $\rm E_{PFN}$ with and without feed-forward control.  The data are shown in \% of RMS divided by the mean. }
\footnotesize
\begin{tabular}{|l|ccccccccc|}\hline\hline
mod.&\hspace{-2.5mm} \#0 &\hspace{-2.5mm} \#1 &\hspace{-2.5mm} \#2
&\hspace{-2.5mm} \#3 &\hspace{-2.5mm} \#4 &\hspace{-2.5mm} \#5
&\hspace{-2.5mm} \#6 &\hspace{-2.5mm} \#7 &\hspace{-2.5mm} \#8 \\\hline off
&\hspace{-2.5mm} .031 &\hspace{-2.5mm} .089 &\hspace{-2.5mm} .107
&\hspace{-2.5mm} .106 &\hspace{-2.5mm} .015 &\hspace{-2.5mm} .064
&\hspace{-2.5mm} .071 &\hspace{-2.5mm} .097 &\hspace{-2.5mm} .033 \\ on
&\hspace{-2.5mm} .011 &\hspace{-2.5mm} .036 &\hspace{-2.5mm} .070
&\hspace{-2.5mm} .070 &\hspace{-2.5mm} .014 &\hspace{-2.5mm} .029
&\hspace{-2.5mm} .045 &\hspace{-2.5mm} .036 &\hspace{-2.5mm} .009\\\hline\hline
\end{tabular}
\label{sigma}
\end{center}
\end{table}
\vspace{-.5cm}
Fig.~\ref{block} shows block diagram of feed-forward deQ controller.
Upper part shows that for the conventional deQ circuit.  Lower part
shows the feed-forward circuit.  dV/dt is obtained by a differential
circuit. Difference signal of dV/dt and the reference is used to
determine the delay of deQ signal by using a voltage control
delay. Gain of the amplifiers are optimized to make $\rm E_{PFN}$
independent from dV/dt.

\begin{figure}[htb]
\centering
\includegraphics*[width=70mm]{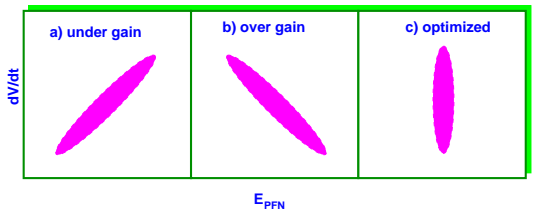}
%\epsfile{file=optimization.eps,width=70mm}
\vspace{-4mm}
\caption{Schematic $\rm dV/dt-E_{PFN}$  plots for three cases;  a) under gain, b) over
gain,  and  c) optimized.}
\label{opti}
\end{figure}
Fig.~\ref{opti} schematically shows $\rm E_{PFN}$-dV/dt correlation in
three cases: a) under gain, b) over gain, and c) optimized. By looking
such plot, we optimized the gain of feed-forward circuit for 9
modulators.

Table~\ref{sigma} shows the stability of $\rm E_{PFN}$ with and
without feed-forward control. The unit is $\sigma E_{PFN}/E_{PFN}$.
Except for modulator \#4, there was significant improvement for the
stability up to factor of 3.

\begin{figure}[htb]
\centering
\includegraphics*[width=60mm,height=52mm]{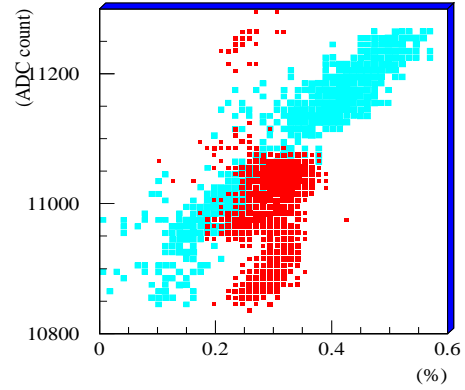}
%\epsfile{file=pjitter.eps,width=60mm,height=52mm}
\vspace{-5mm}
\caption{x and y axe show the momentum jitter, $\rm \Delta P/P$ and amplitude of AC 200V line respectively. }
\label{pjitter}
\end{figure}
Fig.~\ref{pjitter} shows the correlation between momentum jitter, $\rm
\Delta P/P$ and amplitude of AC 200V line.  Momentum jitter was
measured in beam transport line where the dispersion function was
large.  The dispersion function was calculated by a software for beam
dynamics, so called SAD implemented by KEK scientists. Blue and red(or
light and dark) areas show data taken without and with feed-forward
control respectively.  In this plot, we can see a clear dependence of
momentum to the amplitude, but it is compensated by feed-forward deQ.
The momentum jitter was decreased from 0.6\% to 0.2\% peak-to-peak.

\begin{figure}[htb]
\centering
\includegraphics*[width=60mm,height=50mm]{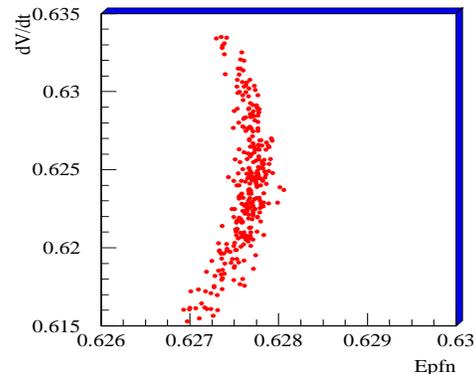}
%\epsfile{file=nonlin.eps,width=60mm,height=50mm}
\vspace{-4mm}
\caption{x and y axe show $\rm E_{PFN}$ and dV/dt respectively.  Data were taken with fully optimized feed-forward controlled deQ. }
\label{nonlin}
\end{figure}

Fig.~\ref{nonlin} shows $\rm E_{PFN} - dV/dt$ curve with feed-forward
control. This strange shape can be explained as follows; although the
feed-forward control should be driven by dV/dt measured at the end of
the charge up curve ($\rm \cot\theta$), dV/dt is actually measured at
the middle of the charge up curve($\rm \cot\theta^\prime$).  The gain
for dV/dt, $\rm G$ is therefore adjusted like $\rm \cot\theta =
G\cot\theta^\prime$ and $\rm G\cot\theta^\prime$ is used instead of
$\rm \cot\theta$.
Due to the sine shape of the charge up curve, the gain $\rm G$ is a
function of deQ timing.  Because deQ timing changes according to
dV/dt, then $\rm G$ is a function of dV/dt. On the other hand, $\rm G$
is a constant in our circuit.  As a result, where dV/dt is
large(small), $\rm G$ is too large (small) and feed-forward control
becomes over(under) gain.

To fix the problem, the gain have to be changed according to input
signal (dV/dt).  By introducing a linearizer to the feed-forward
circuit, the gain can be changed continuously as a function of input
signal.  A modified feed-forward circuit was implemented with three
linearizers.  The test is now in progress.

\section{Phase-lock system}
Because the experimental hall where ATF is placed does not have a good
thermal condition, the klystron gallery temperature is drifted $\rm
\sim 1^\circ C$ in one day. RF phase is also drifted $\rm 3-5^\circ$ of
S-band frequency.

\begin{figure}[htb]
\centering
%\epsfile{file=feedback.eps,width=80mm}
\includegraphics*[width=80mm]{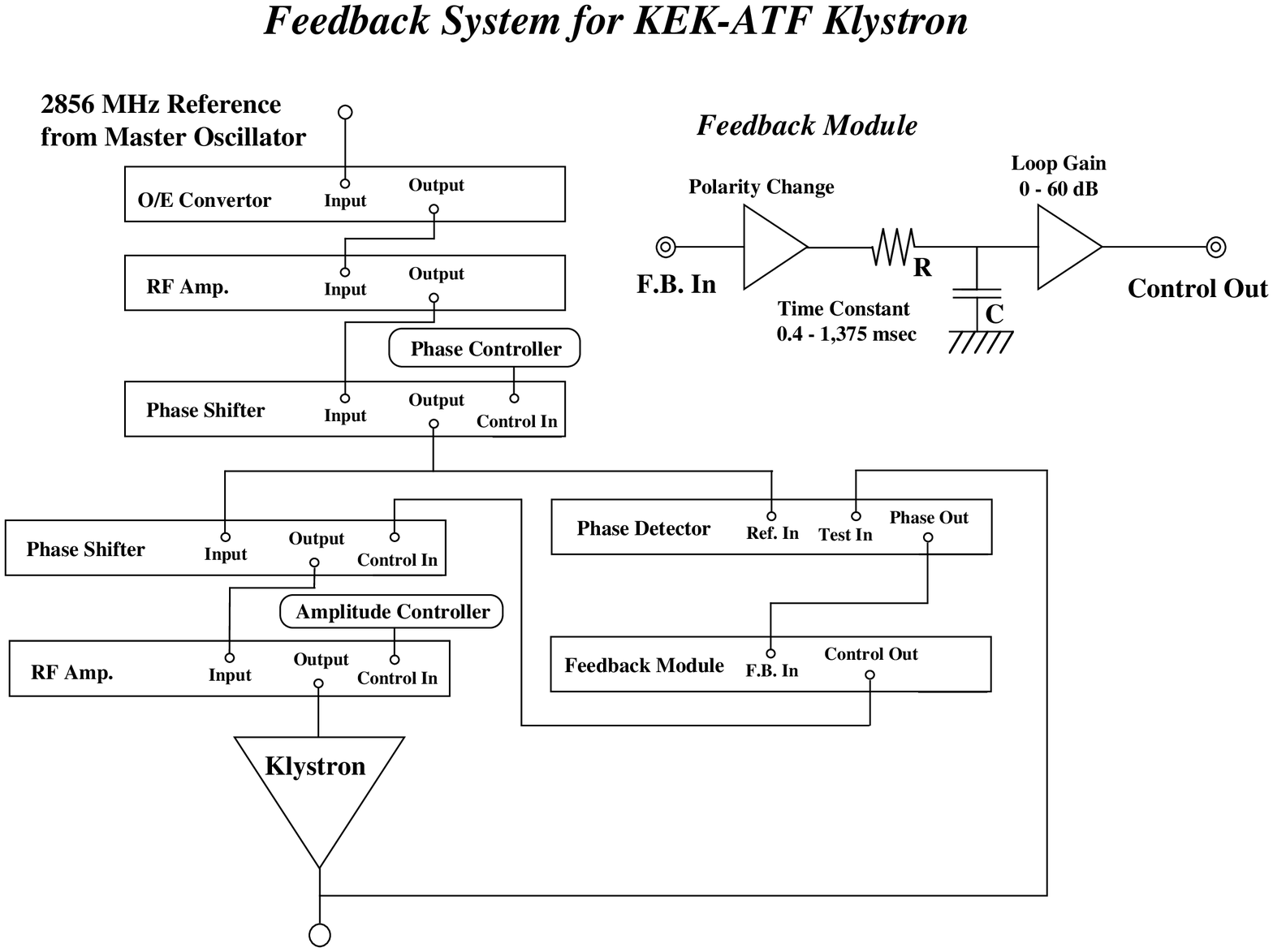}
\vspace{-4mm}
\caption{Block diagram of the phase lock system for the klystron.}
\label{fbblock}
\end{figure}

To stop the RF phase drift, we have introduced a phase lock system.
The block diagram of the system is shown in Fig.~\ref{fbblock}.
RF reference distributed through an optical line is separated into two
lines after the phase shifter. One is for the reference of the phase
detector which observes phase drift at the klystron output.  Another
is for klystron input through another phase shifter controlled by the
phase-lock system to absorb klystron phase drift.  Feed-back module
gains the phase detector output and cuts the high frequency component
to prevent oscillation of the feed-back loop.

The phase-detector is developed by Nihon-koshuha originally for KEKB
RF system. For our purpose, it was modified for pulsed RF by adding
sample hold circuit.  In this module, phase is measured by taking ExOR
of two inputs gained by limiter amplifiers.  Due to this logic, output
phase is independent from input RF amplitude.  Table~\ref{pdspec}
shows the specifications.

\begin{figure}[htb]
\centering
%\epsfile{file=ResultFeedback.eps,width=70mm}
\includegraphics*[width=70mm]{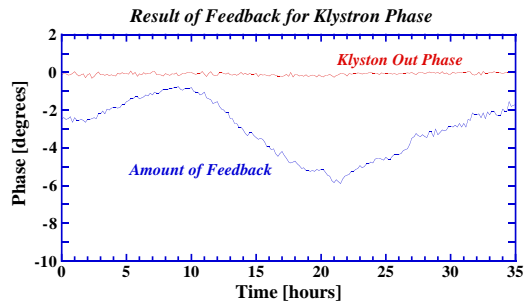}
\vspace{-4mm}
\caption{Stabilized klystron phase and feed-back amount.}
\label{fbresult}
\end{figure}

\begin{table}[htbp]
\begin{center}
\caption{Spec. of the phase detector
( resolution was obtained at reference 0dBm, test $\rm -15\sim0dBm$) }
\begin{tabular}{|l|c|}\hline\hline
item & spec. \\\hline
frequency 		& $\rm 2856\pm 5 MHz$\\
input pulse duration	& 300 ns $\sim$ C.W. \\
reference dynamic range & $\rm -30\sim +5 dBm$\\
test dynamic range & $\rm -10\sim +5 dBm$\\
phase detectable scope & $\rm 540^\circ$\\
phase resolution	   & $\rm\pm 1^\circ$ \\\hline\hline
\end{tabular}
\label{pdspec}
\end{center}
\end{table}
Fig.~\ref{fbresult} shows a result of a test for the phase-lock
system. Klystron RF phase was stabilized within $\rm 0.2^\circ$
peak-to-peak by the phase-lock system with the amount of the feed-back
up to $\rm 6^\circ$.  Accounting for $\rm 1^\circ$ systematic error,
the klystron RF phase was stabilized within $\rm 1^\circ$.
The phase-lock system will be introduced for all klystrons soon.  This
system will suppress the long range energy drift of the S-band linac.

\section{summary}
To suppress the energy instability of the S-band linac, feed-forward
control deQ was introduced.  The momentum jitter was measured at the
beam transport line and was improved from 0.6\% to 0.2\% by the
feed-forward deQ.  For further improvement, the feed-forward circuit
was modified to correct the curve shape of dV/dt-$\rm E_{PFN}$
relation by using linearizers.

For the long term phase drift of klystron RF, a phase-lock system was
examined and stabilized the RF phase within $\rm 1^\circ$. 
The phase-lock system will be introduced for all klystrons soon.

\end{document}